\documentclass[anonymous]{aastex62}

\newcommand{\rep}[1]{{#1}}
\newcommand{\repII}[1]{{#1}}
\newcommand{\new}[1]{{#1}}

\submitjournal{ApJ}

\shorttitle{Clumping in Wolf-Rayet Stars}
\begin{document}

\title{Clumping in the Winds of Wolf-Rayet Stars}

\correspondingauthor{Andr\'e-Nicolas Chen\'e}
\email{andrenicolas.chene@gmail.com}

\author[0000-0002-1115-6559]{Andr\'e-Nicolas Chen\'e}
\affil{Gemini Observatory/NSF’s NOIRLab, 670 N. A`ohoku Place, Hilo, Hawai`i, 96720, USA}

\author[0000-0003-3890-3400]{Nicole St-Louis}
\affiliation{D\'epartement de Physique, Universit\'e de Montr\'eal, C. P. 6128, succ. centre-ville, Montr\'eal (Qc) H3C 3J7, and\\
 Centre de Recherche en Astrophysique du Qu\'ebec, Canada}

\author[0000-0002-4333-9755]{Anthony~F.~J. Moffat}
\affiliation{D\'epartement de Physique, Universit\'e de Montr\'eal, C. P. 6128, succ. centre-ville, Montr\'eal (Qc) H3C 3J7, and\\
 Centre de Recherche en Astrophysique du Qu\'ebec, Canada}

\author{Kenneth G. Gayley}
\affiliation{Department of Physics and Astronomy, University of Iowa, Iowa City, IA 52242, USA}



\begin{abstract}

We attempt to determine the driver for clumping in hot-star winds by extending the measure of the spectral variability level of Galactic Wolf-Rayet stars to by far the hottest known among them, \rep{the WN2 star WR\,2} and the WO2 stars WR\,102 and WR\,142. These \rep{three stars have T$_{\star}$ = 140 kK and 200 kK, the last two being well above the bulk of WR stars with T$_{\star}\sim$ 40 -- 120 kK}. This full temperature range for WR stars is much broader than that of their O-star progenitors  ($\sim$30-50 kK), so is better suited to look for any temperature dependence of wind clumping. We have obtained multiple observations with high signal-to-noise, moderate-resolution spectroscopy in search of small-scale variability in the strong emission lines from the dense winds of these \rep{three extreme stars}, and find a very low-level of variability in both stars. Temperature and terminal velocity are correlated, so faster winds show a lower variability, though this trend goes against any predictions made involving Line Deshadowing Instability (LDI) only\rep{, implying that instabilities intrinsic to LDI are not the main source of wind clumping}. Instead, it could be taken as support for the suggestion that clumps are caused by a sub-surface convection zone (SSCZ) at T $\sim$ 170 kK, since such an SSCZ would have little opportunity to operate under the hydrostatic surface of these hottest WR stars. It is still possible, however, that an SSCZ-related driver could interact with nonlinear line instability effects to enhance or possibly even produce clumps.

\end{abstract}

%editorials, notices --- miscellaneous --- catalogs --- surveys}


\section{Introduction} \label{sec:intro}

The discovery more than 30 years ago of stochastic clumps in the winds of two Wolf-Rayet (WR) stars \citep{Sc75,Mo88} led to the current evidence that most, if not all, hot-star winds are pervaded by hierarchies of thousands of small, randomly varying clumps \citep{Le99,Le08,Ev98,Gr01}. This fundamental discovery has had a profound effect on our understanding of hot stellar winds, perhaps the most important of which is the necessary reduction in the estimated mass-loss rates by a factor 2-5 \citep{Mo94}. This in turn affects the whole evolutionary history of massive stars, where mass-loss plays a crucial role \citep{Ma00,Ek12}.

But the question of the origin of these structures remains uncertain and controversial. On the one hand, clumps can be generated spontaneously in the wind by the inherently unstable nature of radiative line-driving \citep{Lu80,Ow88}, known as Line Deshadowing Instability \citep[LDI, equivalently Line Driven Instability,][]{Ow84,Su13}. Based on this theory, all hot stars should have clumpy winds, and especially the hotter ones, since multiplying the flow time by the growth rate of small-scale line-driven perturbations yields of order $v/v_{th}\sim 100$ in all but the most extreme cases of multiple scattering \citep{Ga95}.% Also, in the absence of another driver, clumps would only occur starting at a given radius \citep[calculated to be 1.1 R$_\ast$, where R$_\ast$ is the hydrostatic stellar radius, by][]{Su13} below which the wind is relatively smooth.

On the other hand, it has been claimed that stochastic wind variability of WR stars may be higher in cooler stars \citep{Mi14}, which would require an explanation separate from the predictions of the LDI theory. Classical WR stars, the late He-burning stage of stars with an initial mass above $\sim$25\,M$_\odot$, as well as WNh stars, which are H-rich very massive main sequence stars, have the strongest stable winds of all hot luminous stars, making them ideal candidates to study  optical wind variations. The work by \citet{Mi14} compiles the measurements of the line profile variability (lpv) amplitude, $\sigma$, of 64 Galactic WR stars. This sample includes WR stars of all subtypes in both the WN and WC sequences, and covers surface temperatures (which correlate well with the wind temperatures) between 40 kK and 140 kK.

If confirmed, this trend could suggest that another mechanism at the origin of the clumps may be at play, possibly even surpassing the role of the inherent LDI instability. One candidate mechanism are the sub-surface convection (SSC) regions associated with a partial ionization zone (PIZ) of iron at T$\sim$170kK in luminous hot stars \citep{Ca09}. \citet{Gr16} has already claimed that a sub-surface convective region is found in all their models of  H-free WR stars in the range 2$-$17 M$_\odot$. When the stochastic motions associated with this convection reach the (radiative) stellar surface, they could generate clumpy structures, altering the driving of mass flux through the critical point, thereby altering the density or the radiative flux at the surface. \rep{If that mechanism is the main drivers of clumps, it implies that slower and cooler winds are clumpier}, since such stars have cooler surfaces and deeper and denser subsurface convection zones. Since smaller, hotter stars have higher escape speeds and faster winds, temperature provides an effective lever for distinguishing these possible sources for clumping.

In this study, we revisit the trend presented by \citet{Mi14} by adding 29 stars and recalculating the $\sigma$ values uniformly for the whole sample. Two of the new stars are WR\,102 and WR\,142, both of subtype WO2; they are by far the hottest among known Galactic WR stars with T$_\ast = 200$\,kK \citep{Sa12}. We also revisit the star WR\,2 (WN2b), the hottest known Galactic WN star \citep[T$_\ast = 141.3$\,kK,][]{Ha06}, for which no clump-related lpv has been found at, or above a level of 1\% of the line intensity \citep{Ch08}, in contrast with all other observed WR stars \citep{Ro92,Le99,Ch11b}. These three hot WR stars at the extreme limit offer an opportunity to test if the amplitude of lpv depends significantly on the wind temperature and, consequently, if the origin of the hot-stellar wind clumps involves a mechanism like subsurface convection.  In Section 2, we present our new observations and describe how we carried out the data reduction. Section 3 relates how we carried out the uniform evaluation of the level of spectroscopic variability and Section 4 discusses the special case of the three hottest stars in our sample. We discuss our results in Sections 5 and 6 and conclude in Section 7.

%As this is already hotter than the PIZ, these stars give us the opportunity to focus exclusively on the variations promoted by the line-driven instability, as the stellar surface should be relatively unperturbed and unmagnetized by convection.

% If we find no or little variability in the wind of the three stars, supporting the claimed trend in stochastic wind variability is skewed toward cooler WR stars \citep{Mi14}, this will confirm that the primary origin of wind clumping in hot-star winds lies in subsurface convection (B). Otherwise, this will confirm the wind-instability origin (A).

\section{Observations and reduction}
The data used for most of the stars in this study have already been published in \citet{St09}, \citet{Ch11b} and \citet{Gr16} (see Section\,\ref{sigma}). However, new spectra were observed for WR\,2, WR\,102 and WR\,142 using the GMOS spectrographs at Gemini-North and Gemini-South (see Table\,\ref{obsdetails} for details on the observing campaigns).

\begin{table}[h!]%\label{obsdetails}
\renewcommand{\thetable}{\arabic{table}}
\centering
\caption{Observing campaigns description} \label{obsdetails}
\begin{tabular}{lccccl}
\tablewidth{0pt}
\hline
\hline
Star name & Wavelength  & Exposure time & S/N (continuum) & number of & Program ID\\
          & coverage    &  per frame    & per frame      & spectra & \\
\hline
WR\,2   & 3970 -- 5435\,\AA & 120s/250s & $\sim$125 & 68 & GN-2012B-Q-115 \\
WR\,102 & 4380 -- 6000\,\AA & 120s & $\sim$50 & 82 & GS-2017B-Q-62  \\
WR\,142 & 4370 -- 6100\,\AA & 120s & $\sim$100 & 74 &  GN-2017B-Q-6  \\
\hline
\end{tabular}
\end{table}

\begin{figure}[htbp]
%\epsscale{0.9}
\plotone{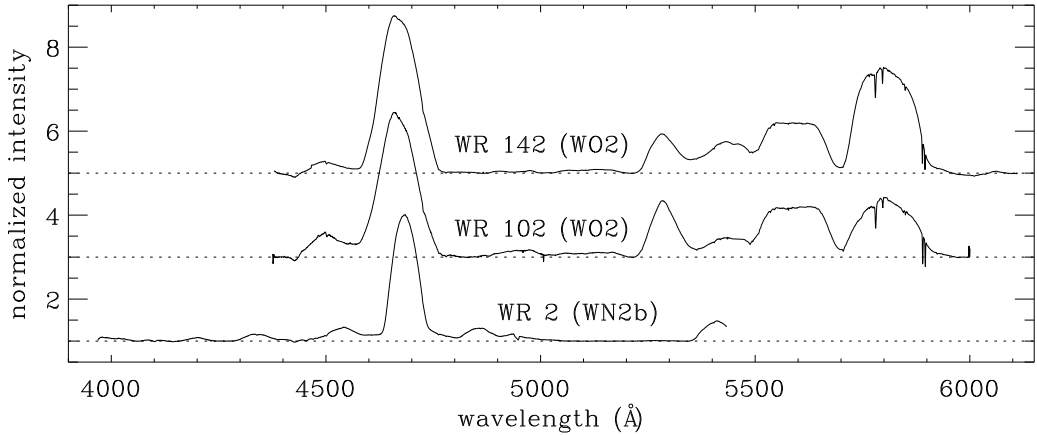}
 \caption{Average of all the Gemini spectra observed during our campaigns for the three stars WR\,2 (WN2b), WR\,102 (WO2) and WR\,142 (WO2). The spectra are arbitrarily shifted vertically from one star to the next for clarity. The dashed horizontal line for each star marks the rectified continuum level.}
  \label{mean}
\end{figure}

For these observations, we reached a 1.5\,\AA\,spectral resolution (R~$\sim$~3400), which is ideal for resolving the typical small-scale variations due to clumps seen in all other well-observed WR stars. The rapid cadence of observation ($\sim$~140s between each spectrum) allows us to prevent motion blur of the variations due to rapidly accelerating clumps.

The bias subtraction, flat-fielding, spectrum extraction, sky subtraction and wavelength calibration of all spectra were executed in the usual way using the Gemini {\sc iraf}\footnote{{\sc iraf} was distributed by the National Optical Astronomy Observatories (NOAO), which was operated by the Association of Universities for Research in Astronomy, Inc. (AURA) under cooperative agreement with the National Science Foundation (NSF).} package. 

Special care was taken for the normalization of the spectra. First, a mean was made for each of the series that were taken during the same night. Then each spectrum of a given night was divided by the night mean and the ratio fitted with a low-order Legendre polynomial (between 4$^{\rm{th}}$ and 8$^{\rm{th}}$ order). The original individual spectra were divided by this fit and were therefore at the same level as the night mean. Once this procedure was completed for each night, the night means were then put at the same level by using the same procedure as described above, and a global, high quality mean was produced. This final mean spectrum was used to put all individual spectra of the entire run at the same level and was also fitted in selected pseudo-continuum regions, defined as wavelength regions where large emission lines do not dominate. These regions lie on the straight horizontal lines in Figure\,\ref{mean}. This fit was applied to all individual spectra, so they become normalized by the continuum in a uniform way. The error on the continuum normalization measured as the standard deviation of individual spectra around the continuum function is typically 0.5\% per pixel.

%More specifically, these regions are~: 4047.5 -- 4053.0\,\AA, 4096.8 -- 4099.3\,\AA, 4131.0 -- 4135.5\,\AA, 4251.0 -- 4280\,\AA, 4393.5 -- 4401.0\,\AA\,and 5058.0 -- 5227.0\,\AA. 

\section{Recalculating the $\sigma$ values}\label{sigma}

\subsection{Determining $\sigma$}

Since a few cases presented by \citet{Mi14} were moderate outlyers, and because we are merging values published in separate studies, we recalculated the $\sigma$ values for all the observed WR stars in a uniform way, with a much better control of the systematics.

We worked from the reduced and normalized spectra published in \citet{St09}, \citet{Ch11b} and \citet{Gr16}, and observed with the 1.6m telescope at the Observatoire du Mont M\'egantic, the 1.5m telescope at the Cerro-Tololo International Observatory and the Gemini Observatory\footnote{Program IDs : GS-2008B-Q-87, GS-2010B-Q-58, GN-2010B-Q-78, GN-2010B-Q-119, GS-2014A-Q-42, GS-2014A-Q-73 and GN-2018B-Q-404}.

These sources have a total of 4 to 5 observed spectra per star, which allows one to determine the statistical significance of the observed changes using the Temporal Variance Spectrum (TVS) formalism as introduced in \citet{Fu96}. The square root of the TVS, which divided by the detection threshold becomes $\Sigma$, is proportional to the amplitude of variability at each wavelength, and is compared with a detection threshold. For stars with a significant detection of lpv, we calculate the $\sigma$ value, extracted from the $\sigma$ spectrum, as introduced by \citet{St09}. The $\sigma$ spectrum differs from the $\Sigma$ spectrum, in the sense that instead of comparing the variance at a given pixel with the noise in the neighboring continuum, it compares the variance of a given pixel with the noise in that same pixel. The advantage over $\Sigma$ is that it gives the amplitude of the variability relative to the intensity of the line over which it was observed. In other words, it gives a measure of how high in percentage of the line intensity the amplitude of the lpv is. 

\rep{The equation for $\sigma$ diverges when it gets too close to the continuum, as it has no meaning outside of emission lines. That is why we can only determine $\sigma$ around the center of the lines, avoiding the low intensity part of the wings. It is for that reason, and for another reason that will be discussed shortly below, that we determine the $\sigma$ value for a given star by calculating the median of the $\sigma$ spectrum within one FWHM of the emission line around line center.

Our results are presented in Table\,\ref{tab:WC} for WC and WO stars, in Table \ref{tab:WN} for WN stars, in Table \ref{tab:WNH} for WNh stars and in Table \ref{tab:WNC} for WN/WC stars.}

\subsection{Error sources and uncertainties}

\rep{The main challenge of this study is that we are using archival data originally planned for a different purpose. The spectra were designed for the search of large-scale lpvs that are easy to detect and measure when present on the top of virtually any mildly strong spectral emission line. This leads to a certain number of error sources that need tailored mitigation.

One main error source, which we can easily characterise, is the photon noise of the spectra. Indeed, the equation for the $\sigma$ spectrum does not differentiate real variations from the noise. Consequently, $\sigma$ is artificially high and diverges from the real lpv amplitude when the variation level is close to the noise level. With a signal-to-noise ratio (S/N) for the spectra between 70 and 100 in the continuum, the $\sigma$ spectrum diverges when flux gets as low as $\sim$2.5 times the continuum or lower (though this value may be higher when the S/N is lower). This the other of the two reasons (see above) why the $\sigma$ value is calculated only within one FWHM of the emission line, since doing so respects this conditions in most of the cases. 

Another issue occurs with very strong emissions lines that can't be normalized with an accuracy better than 1\% due to a lack of continuum over a large wavelength range. In those cases, the lpvs can be affected by a low frequency variation that increases artificially the $\sigma$ value. This problem varies with the line intensity, the separation between the regions of continuum to the line, and the complexity of the continuum shape (dominated by the spectrograph's illumination profile). We avoid those lines as much as we can.

In some cases, spectra continuum is variable due to an external source (e.g., a potential binary companion or stray light in the instrument). For those stars, the lines were normalized to a uniform intensity, assuming that the clump variability would be sufficiently unaffected by variations taking place over a broad wavelength range.

Finally, time sampling also contributes to the uncertainties. Indeed, c}lumps are stochastic, and may translate into any number of lpvs anywhere on the spectral lines at any time. With a proper monitoring over a few hours repeated during a certain number of nights, it is possible to monitor well the lpv over the whole line. However, with a limited number of spectra, like the 4-5 spectra we are working from, we may just sample some of the lpv and miss the complete picture. Consequently, the $\sigma$ spectrum may end up lower and more skewed than it would be with better sampling. One remedy to these setbacks would be to define the $\sigma$ value as the maximum value of the $\sigma$ spectrum. But because $\sigma$ may be overly influenced by noise, and because there would be no way to assess a statistically significant uncertainty on the value, such an approach would introduce additional systematics, and is non recommended. Instead, we opt for the median value\rep{, which is more robust. 

We define the error on $\sigma$ as the standard deviation of the $\sigma$ spectrum within the wavelength range used to calculate $\sigma$. This may not correspond to the standard definition of error bars, but it does represent well the effect of our limited time sampling. If there are no better lines than those affected by the continuum normalization issue, we add 1\% to the final error budget on $\sigma$. Finally, if only faint spectral lines were available for a given star, and the noise level was close to the amplitude of the lpv, we considered the lpv as not detected, and used the detection threshold defined by the TVS as the maximum value possible. The errors are typically around 1\%, or less when the S/N atop the line is high. The error is a few tenths of a percent for WR\,2 and WR\,142, as $\sim$70 spectra where used instead of 4-5. The same would apply to WR\,102 if variability was detected.}

\subsection{$\sigma$ dependence on spectral line}\label{linechoice}

Ideally, the same spectral line would be used for all the stars to measure $\sigma$. And the best candidates would be He{\sc ii}$\lambda$4686 for WN stars and C{\sc iii}$\lambda$5696 for WC stars, as they are both singlets originating at large radii in the wind. However, this was not possible for two reasons. The first one is the different wavelength range covered during the multiple observing campaigns that produced the data we are using. The second, inevitable, is the dependence of the spectral lines intensity with the temperature (or spectral type).

Our analysis is therefore based on the assumption that the $\sigma$ calculated for a given line is statistically equivalent to the value calculated for any other line. This assumption was verified by \citet{Ch11b}, when different spectral lines than for \citet{St09} were used. \new{Indeed, the results for WR\,111 and WR\,115 were the same regardless of which spectral lines were studied}. \citet{Ch10} also plot the $\sigma$ spectrum for WR\,1 for 9 lines which did not all give the exact same value, but differences were below 1\% (for lines that are not affected by any of the issues mentioned above), comparable to the error bars. \rep{On the other hand, \citep{Le00} observations of WR\,135 show that the C{\sc iv}$\lambda$ 5808 doublet region gives in that case a systematically lower $\sigma$ than other lines.

To make sure our assumption is valid, we analysed as many spectral line per star as we could use, making sure to use only the best line, as free as possible from systematic issues. Looking at the Tables\,\ref{tab:WC} to \ref{tab:WNC}, we find that the difference in $\sigma$ between lines is always within the error, but we few exceptions. In some cases, one of the line intensity is close to the limit at which noise contributes to $\sigma$ significantly, potentially causing the associated $\sigma$ to be over evaluated. More importantly, the C{\sc iv}$\lambda$ 5808 doublet region is effectively systematically less variable for WC7 and WC8 stars. We suspect that at their temperature, the contribution of the two components of the line's doublet nature blends the variations and falsely create lower amplitude residuals. We therefore avoid using that line for those stars.}

\section{Line profile variability of WR\,2, WR\,102 and WR\,142}\label{Spec}

\begin{figure}[htbp]
\epsscale{0.7}
%\plottwo{WR2_montage.eps}{WR2_sigi.eps}
\plotone{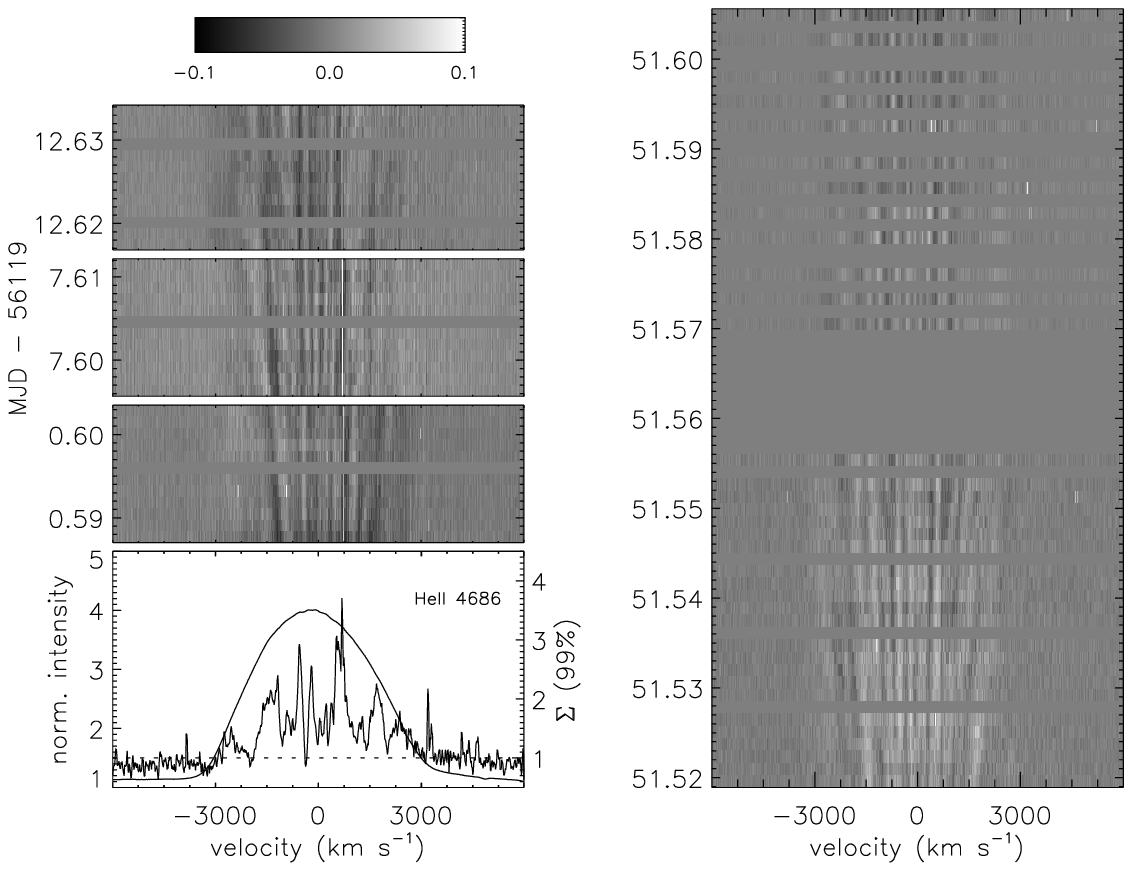}
\epsscale{0.4}
\plotone{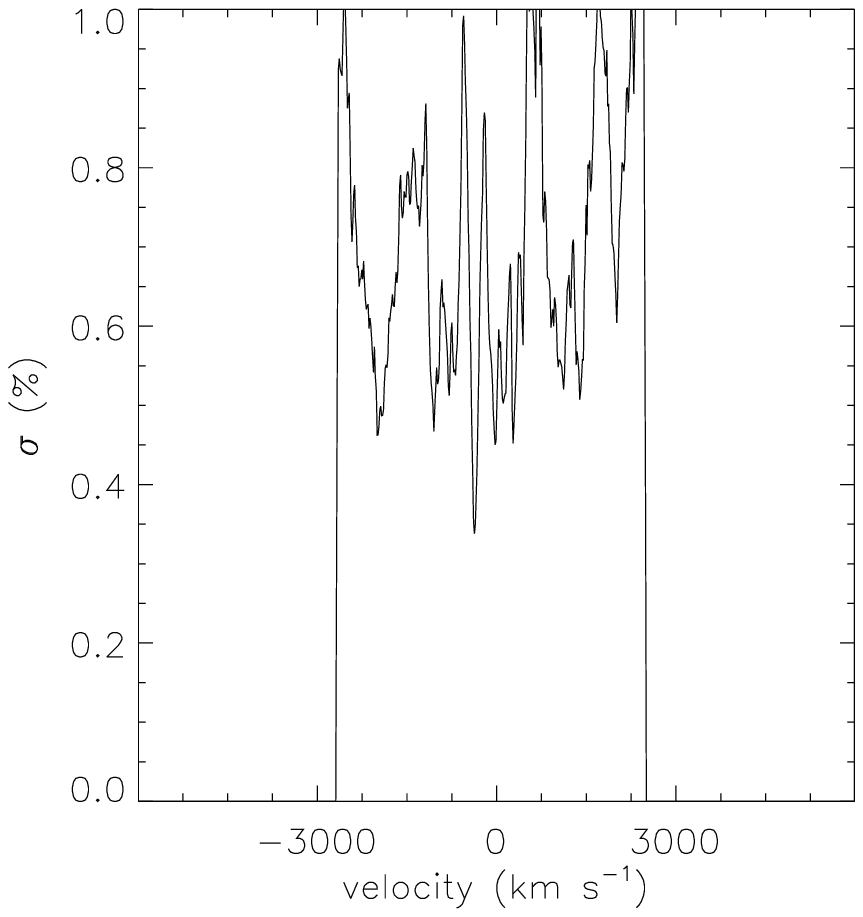}
 \caption{Montage of our GMOS spectra showing the spectral variability of the He\,{\sc ii}$\lambda$4686 line of WR\,2. {\it Left:} The bottom panel shows the mean and the $\Sigma$ spectrum showing the level of significance of the line profile variability. The top panels show the time-resolved greyscale plot of the residuals as a function of time after subtracting the mean spectrum for the first three nights. {\it Middle:} Time-resolved greyscale plot for the forth night. {\it Right:} $\sigma$ spectrum indicating the fraction of the line flux that is variable.}\label{mont2}
\end{figure}

\begin{figure}[htbp]
%\epsscale{0.9}
\plotone{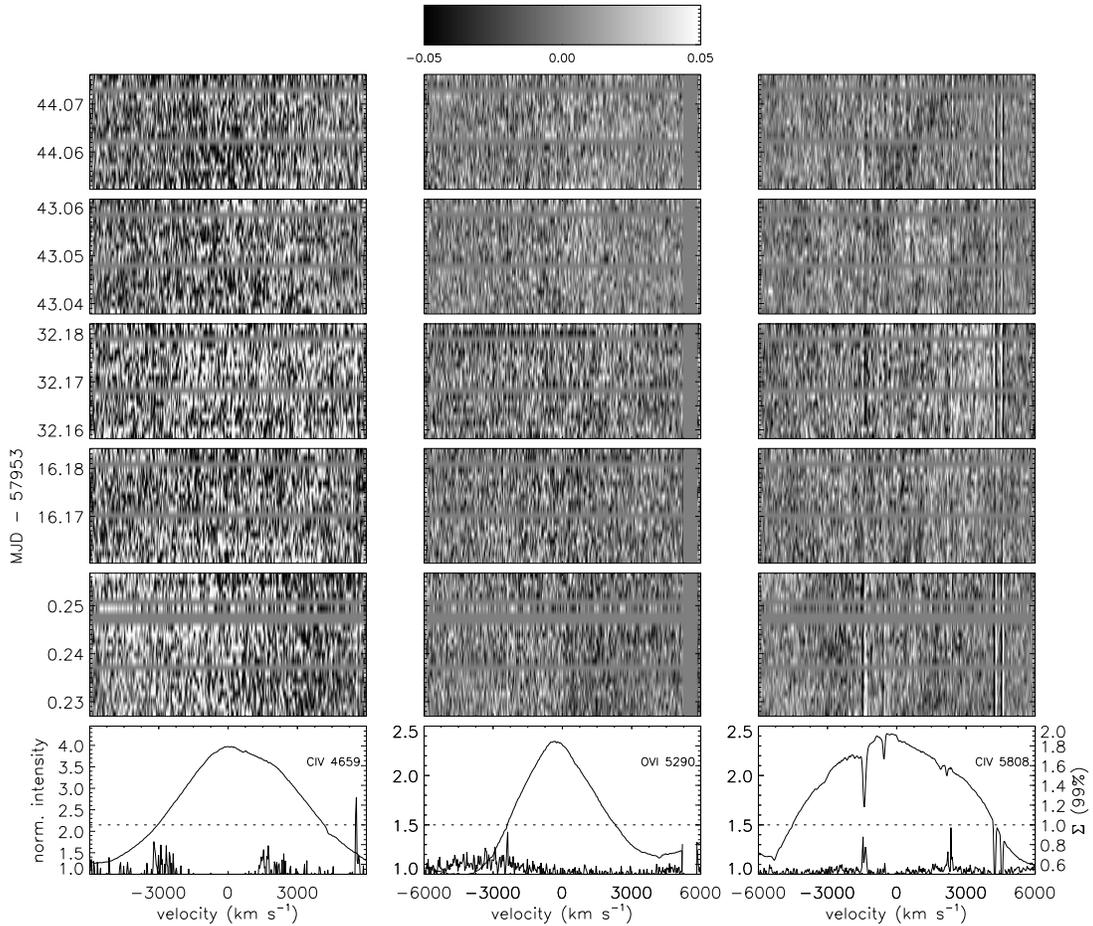}
 \caption{Montage of our GMOS spectra showing the spectral variability of \rep{the C\,{\sc iii}/{\sc iv}$\lambda$4659 multiplex, the O\,{\sc vi}$\lambda$5290 line, and the C\,{\sc iv}$\lambda$5808 doublet region} of WR\,102. {\it Bottom panels:} Mean and $\Sigma$ spectrum showing the level of significance of the line profile variability.}\label{mont102}
\end{figure}

\begin{figure}
%\epsscale{0.9}
\plotone{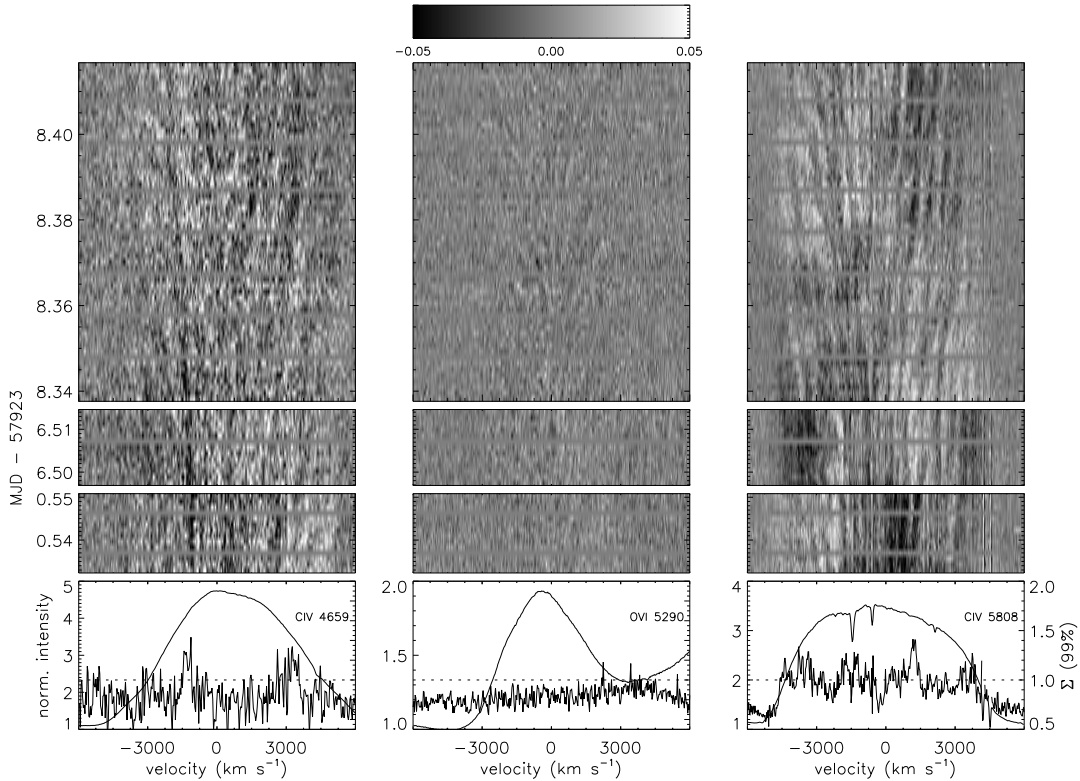}
 \caption{Same as Figure\,\ref{mont102}, but for WR\,142.}\label{mont142}
\end{figure}

In this section, we present the observed lpvs of the 3 WR stars that we are adding to the sample, WR\,2, WR\,102 and WR\,142. The lpvs are illustrated in Figures\,\ref{mont2}, \ref{mont102} and \ref{mont142}, respectively. The residuals created by subtracting the mean spectrum from individual spectra are shown in the grayscale plots. The $\Sigma$ spectrum is presented in the bottom panel. 

\subsection{WR\,2}

The wavelength range only allows for the monitoring of the He\,{\sc ii}$\lambda$4686 line changes. All other lines observed have a lower intensity, and our spectra at those wavelengths do not reach a S/N high enough for a significant detection of low amplitude lpv.

Each series observed during the first three epochs (plotted on the left side of Figure\,\ref{mont2}) contains 11 spectra and covers $\sim$~45 min. Even in these short datasets, one can clearly see rapidly moving sub-peaks on top of the line, indicating that clumps are present in WR\,2's wind. However the contiguous time-span is not long enough to monitor individual sub-peaks from the moment they appear to the moment they dissipate. Luckily, we obtained a fourth series that covers a total of $\sim$~2.6 h (middle of Figure\,\ref{mont2}). Due to adverse weather conditions, the series was cut into two 1.2 h series, the second of which was obtained using an exposure time of 250 sec (slightly more than twice as long as for the previous spectra), to compensate for the absorption caused by passing clouds.

The TVS \new{(calculated for the four nights all together)} shows that the He\,{\sc ii}$\lambda$4686 line is variable over the whole wavelength range covered by the line. The detected lpv is well above the threshold where the variations are significant at the 99\% confidence level, as defined by the TVS. All the sub-peaks move towards the wings of the line and never cross the central wavelength, as expected for systematic clump motion following the general wind expansion. Hence, clumps are definitely detected in the wind of WR\,2. 

One interesting result is that the variability extends well above the terminal velocity found by \citet{Sh14} of $v_\infty=1800 $\,km\,s$^{-1}$. A model of WR\,2 briefly presented in \citet{Ch19}, yielded a terminal velocity of $v_\infty=3200 $\,km\,s$^{-1}$, closer to this value \rep{and matching older spectral analysis from \citet{Sc89}}. We therefore suggest that the wind of WR\,2 reaches a much higher terminal velocity than previously thought.

A measure of the standard deviation of the intensity variations at each wavelength \citep[as presented by][for other WR stars]{St09} indicates that the sub-peaks have an average rms of 0.6\% of the line intensity, which is half the amplitude observed in the WR winds showing the smallest clump-like variations in the study of \citet{Ro92}. This explains why clumping was not detected in WR\,2 in previous datasets with lower S/N. %Interestingly, assuming a mass of 9\,M$_\odot$ for WR\,2 \citep{Ha06}, a line variability level of 0.5\% fits well within the correlation between stellar mass and $\sigma$ established by \citet{Gr16}.

%The search for periodicity in the changes over time of the different moments of the He\,{\sc ii}$\lambda$4686 line (the strongest line in our spectral range), such as the equivalent width, the RV, the skewness and the kurtosis (each of which varies with an amplitude equal to or lower than 0.1\% over the whole run, and equal to or lower than 0.05\% each night), failed to reveal any stable period. We searched between periods of 10 min and 20 days, using a step in frequency of 0.0002~d$^{-1}$. We also searched for periodicity in the scalar value $\sigma_{\rm res}$, i.e., the standard deviation of the residuals obtained from the subtraction of two spectra observed with a given time, as defined by \citet{Ch10}. This allowed us to verify if there is any timescale covered by the observations at which the spectral features repeat. In both cases, we found that no periodic pattern above a threshold of 2-$\sigma$ can be seen in the way the sub-peaks appear, move and dissipate in the spectrum of WR\,2. 

\subsection{WR\,102}
We monitored the lpv of three of the WR\,102 strongest emission lines, i.e., \rep{the C\,{\sc iii}/{\sc iv}$\lambda$4659 multiplex, the O\,{\sc vi}$\lambda$5290 line,} and the C\,{\sc iv}$\lambda$5808 doublet region. Our 5 sequences of $\sim$~45\,mins each obtained during different nights do not seem to reveal any significant variability (see Figure\,\ref{mont102}). The TVS is far below the threshold where the variations are significant at the 99\% confidence level, even on top of the strong C\,{\sc iii}/{\sc iv}$\lambda$4659 multiplex.

One could be more or less convinced by looking at Figure\,\ref{mont102} that some consistent structures could be seen traveling on top of the lines. WR\,102 (V$\sim$14.1\,mag) is fainter than WR\,2 and WR\,142 and was observed under the worst weather conditions; thus, it is possible that its S/N was insufficient for detecting clumps. We therefore use the S/N on top of the C\,{\sc iii}/{\sc iv}$\lambda$4659 multiplex to determine the maximum amplitude limit clumps should have (in percent of the line's intensity) to be at the same level as the noise in our data. This value is 0.14\% of the line intensity.

\subsection{WR\,142}
We monitored the lpvs of the same three lines of WR\,142 as for WR\,102. During the first two 45\,min long runs, we could see faint subpeaks moving on top of all the emission lines. We therefore used all the remaining observing time of our program to obtain a 2h-long continuous monitoring. The clumps are barely visible and the trajectory of the sub-peaks be traced only with difficulty. But the TVS confirms that we get a significant level of variability that is of comparable amplitude as the noise itself. We therefore interpret this as a significant detection of clumps with an amplitude equal to the noise level on top of the C\,{\sc iii}/{\sc iv}$\lambda$4659 multiplex, which converts to an amplitude of 0.12\% of the line intensity.

\section{lpv dependence on surface temperature} 

\begin{figure}\label{SigvsT}
%\epsscale{0.9}
\plotone{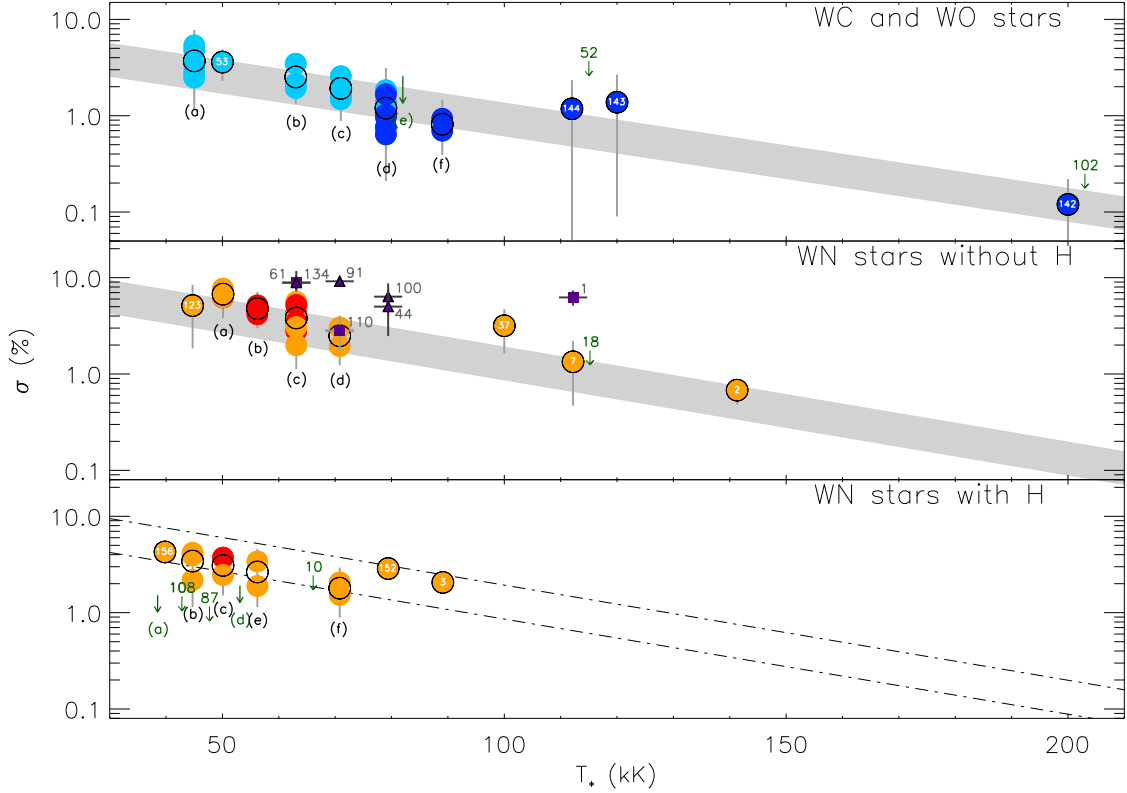}
 \caption{Top panel: $\sigma$ vs T$_{\star}$ for WC and WO stars. Points in light blue are $\sigma$ values obtained from C{\sc iii}$\lambda$5696, while those obtained from C{\sc iv}$\lambda$4659 or 5808 are marked with points in dark blue. Note the logarithmic scale for $\sigma$. For temperatures with only one WR star, the $\sigma$ value is marked by a blue circle with the WR ID annotated. When there is more than one WR star with the same temperature, individual points are not identified, but instead the group of points is given a label from (a) to (e) (see Table\,\ref{tab:WC} for the list of stars within each group). The average in each case is marked by a black empty circle. When clumps were not detected, an upper limit is determined based on the quality of the data. Those cases are marked by downward arrows, slightly offset in temperature by 3kK for better visibility. A tentative linear relation is plotted with a gray bar whose width corresponds to the rms scatter about the mean relation. 
 Middle panel: Same as top panel, but for WN stars without hydrogen. Orange points are $\sigma$ values obtained from He{\sc ii}$\lambda$4686, while those obtained from He{\sc ii}$\lambda$5412 are marked with red points. The stars under the labels (a) to (d) are given in Table\,\ref{tab:WN}. The purple points mark cases where sources other than clumping are the prime cause of the line profile variability. Squares are for CIRs with determined periods, and triangles for CIR-candidates.
 Bottom panel: Same as top panel, but for WN stars with hydrogen. The stars under the labels (a) to (g) are given in Table\,\ref{tab:WNH}. The dash-dotted lines mark the same linear relation as found for the WN stars without H, for comparison.}
\end{figure}

New improved $\sigma$ values for 90 WR stars, now including three extremely hot WR stars with $\sigma$ values determined to 0.1\% accuracy (due to the larger number of high-S/N spectra used), allow us to re-examine the possible relationship between the amplitude of the variability due to clumping and the temperature at the hydrostatic stellar surface (Figure\,\ref{SigvsT}), as first presented by \citet{Mi14}. 

%correlations  WC and WO stars
%    -0.853226      0.00000
%pentes  WC and WO stars
%     0.910624  -0.00916006
%correlations  WN stars without H
%    -0.799744  1.35899e-05
%pentes  WN stars without H
%      1.26987   -0.0101724
 
The top panel of Figure\,\ref{SigvsT} presents our results in the form of log($\sigma$) as a function of T$_{\star}$ (at the hydrostatic stellar surface, obtained by spectroscopic modeling from \citet{Ha06} and \citet{Sa12}) for stars of types limited to WC and WO. \rep{The $\sigma$ values obtained from different lines are plotted in different shades of blue.} Overall, there is a significant decline of the lpv amplitude as a function of T$_\ast$. Assuming a linear association between $\log(\sigma)$ and T$_{\star}$, the fitted slope is $-(9.0 \pm 2.0) \times10^{-6}$ K$^{-1}$. In this case, the Pearson correlation coefficient $r=-0.85$ and the permutation test gives a $p$-value essentially equal to 0, which supports the hypothesis of a linear association. The dispersion of the $\sigma$ values at a given temperature may appear high, and to be consistent with a slope, WR\,143 and WR\,144 should have clumps about half the amplitude measured. But the overall rms dispersion of the data is comparable to the rms error on the $\sigma$ values, and the observations of the spectral lines for both WR\,143 and WR\,144 were affected by the continuum normalization issues. Indeed, the spectrum of those two WC\,4 stars consists of many weak lines with few very intense lines that are tens of times the intensity of the continuum. Because our observing strategy was not well prepared for that, the best we could do was to get a rough estimate of the $\sigma$ value from one line that offered the best compromise. A dedicated project to those 2 stars, and to WR\,52 which was not observed with sufficient S/N would be necessary to verify if the clumps reach an amplitude of 1\% or less in WC winds with temperatures around 100 kK. Nevertheless, whether the trend is an exponential relationship or a more complex relationship, the lpv in the winds of the cool WCL stars clearly has a significantly higher amplitude, and the lpv in the wind of the hot WO stars has a significantly lower amplitude than the average.

The middle panel shows the same as the top panel, but only for the WN stars that have no H lines in their spectrum. \rep{The $\sigma$ values are plotted in orange or red depending on the line they were calculated from.} The result is slightly more complex than for the WC/O stars. Some stars, like WR\,1, 6, 110 and 134 \citep[][respectively]{Ch10,Mo97,Mc94,Ch11a}, display periodic, large amplitude lpv that are due to the presence of corotating interaction regions (CIRs) in their winds. Those stars are plotted with a small square symbol in Figure\,\ref{SigvsT}. There are other candidates for CIRs, such as stars WR\,44, 58 and 61 (plotted with a small triangle), and only a dedicated monitoring campaign could reveal if there is indeed a period for the lpv, thus disclosing their CIR nature. After excluding WN stars with CIR or CIR-like lpvs, we are left with a trend comparable to that of the WC/O stars (that generally tend not to reveal CIRs), with a slope of $-(10.0 \pm 2.0) \times10^{-6}$ K$^{-1}$. The Pearson correlation coefficient $r=-0.80$ and the permutation test give a $p$-value essentially equal to 0. The slopes for both the WN stars without H and the WC/O stars are essentially the same within their errors, but the WN-star lpvs have generally a higher amplitude. Once again, at a temperature around 100 kK, the stars WR\,7 and WR\,37 seem to deviate from the correlation. Both stars show large  lpvs which could be labeled as CIR-like. In that case, either the stochastic lpv is larger at 100 kK, or both stars do have CIRs in their winds. A dedicated project to search for periods in those two stars, and to also re-observe WR\,18 for which we could only determine an upper limit, would help settle this question.

The bottom panel shows only the WN stars with H lines in their spectra. These stars appear to follow a similar trend as the WN stars without H \new{(here plotted with two dot-dashed lines)}, although possibly at a slightly lower level overall.  %[could verify where the WNL stars fall in that bunch. TM: No need, since most WNL stars have H anyway.]

Note here that an equally good relation is obtained using a log-log plot. In such case, the slopes are $-2.3\pm0.2$ for WC/O stars $-2.03\pm0.$2 for WN stars without H. The Pearson correlation coefficient are $r=-0.85$ and $r=-0.79$, respectively, and the $p$-values are essentially equal to 0.

\rep{Finally, one could wonder if the trend with temperature could be impacted by not using the same spectral lines to calculate $\sigma$ for all the stars of the same spectral type. As mentioned in Section\,\ref{linechoice}, there is a difference between the $\sigma$ values obtained using C{\sc iii} and C{\sc iv} lines for the WC7 and WC8 stars. Yet that difference is merely 1\% at most, while the trend we present shows a deviation bigger than one order of magnitude between the coolest to the hottest stars $\sigma$ values.}

\section{Interpretation}

If the lpv amplitude observed in WR winds is correlated with the surface temperature of the star, what does it tell us about the origin of the clumps? \new{Unfortunately, there is no way for the moment to know precisely where exactly lpvs come from. Is one bump associated with one clump, or a group of clumps? Is the lpv pattern on one spectrum a snapshot of the density distribution of the clumps in the wind, or are they dominated by clumps that are the densest and/or the biggest? Consequently, we cannot easily link the $\sigma$ value, which is a measure of the amplitude of the lpvs, to a specific physical parameter of the clumps. On the other hand, it is safe to assume that it is representative of a certain level of ``clumpiness'' in the wind. One possible objection to this assumption is that the amplitude of the variations in the line profile could be dominated by narrow\rep{ lpvs, but with high amplitude}, while wider and \rep{low amplitude ones} would not contribute as much, even with the same total flux. Yet, there are two reasons this objection does not apply. The first comes from the way we calculate $\sigma$; a narrow and \rep{high amplitude lpv} would give a comparable value to that of a wider and \rep{lower amplitude lpv}, once averaged over the wavelength range of the line and would only give larger error bars. The second reason, which is even stronger, is that in fact \rep{lpvs with high amplitude} tend also to be wider, while \rep{lower amplitude lpvs} tend to be narrowed, as illustrated in \citet{Ch11b}.}

{\it A priori}, \new{the dependence of the level of clumpiness, represented by the lpv's amplitude,} is in contradiction with concluding that the LDI is the sole origin of clumps. But does that disqualify it totally? In this section, we explore the possible ways to explain the observed trend, as there are a few competing ideas that can possibly explain a reduction in the variability of clumping at higher (hydrostatic surface) temperature. 

\subsection{Reduced LDI with temperature}

\rep{Starting with the density, $\rho=\dot{M}/(4\pi R_\ast^2 v_\infty)$, where $\dot{M}$ is the mass-loss rate, and $v_\infty$ the wind's terminal velocity, if we take $R_\ast$ \repII{as the hydrostatic radius}, we can rewrite:
\begin{equation}\label{rho}
    \rho=\dot{M}\sigma T_\ast^4/(L v_\infty),
\end{equation}
where $T_*$ is the stellar temperature \repII{at the hydrostatic surface}, $L$ is the luminosity and $\sigma$ the Stefan-Boltzmann constant. Also, starting with the optical depth $\tau_0=n_0 R_\ast \sigma_T$, where $\sigma_T$ the Thompson's constant, we can develop $n_0$, the number density and get:
\begin{equation}\label{tau}
    \tau_0=\frac{\dot{M} R_\ast \sigma_T \sigma T^4}{L v_\infty \mu_e m_H},
\end{equation}
where $m_H$ is the mass of the hydrogen atom, and $\mu_e=2$ for pure once ionised helium. \repII{We consider only the hydrostatic surface, because it is where clumps are expected to be formed, and where the wind is anchored.} Following the equations\,\ref{rho} and \ref{tau}, a} hotter star will be more compact, with a denser, more optically thick wind \rep{where the continuum is emitted, every other stellar parameters} being kept the same. Optical depth effects can reduce the level of LDI: indeed, the linear growth rate is inversely proportional to the effective optical depth (in the sense of the number of overlapping optically thick lines over a terminal speed). The usual growth rate is the ratio of the terminal speed to the ion thermal speed per flow time, so if one divides that ratio by the number of thick lines per terminal speed, one finds that the number of $e-folds$ of the instability is of the order of the ratio of $c$ to the ion Doppler width (so perhaps 30,000) divided by the number of optically thick lines over the entire spectrum. So to completely suppress the LDI, it would require many thousands of optically thick lines. The effective optical depth also gives the ratio of the wind momentum flux to $L/c$, which would need to be many factors of ten in order to kill the LDI. But it is (barely) possible to do this with very dense winds close to the Eddington limit as in WR stars. In Figure\,\ref{Sigma_vs_efold}, we plot the $\sigma$ value as a function of the $e-folds$ \citep[from ][]{Ga95}:

\begin{figure}
    \centering
    \includegraphics{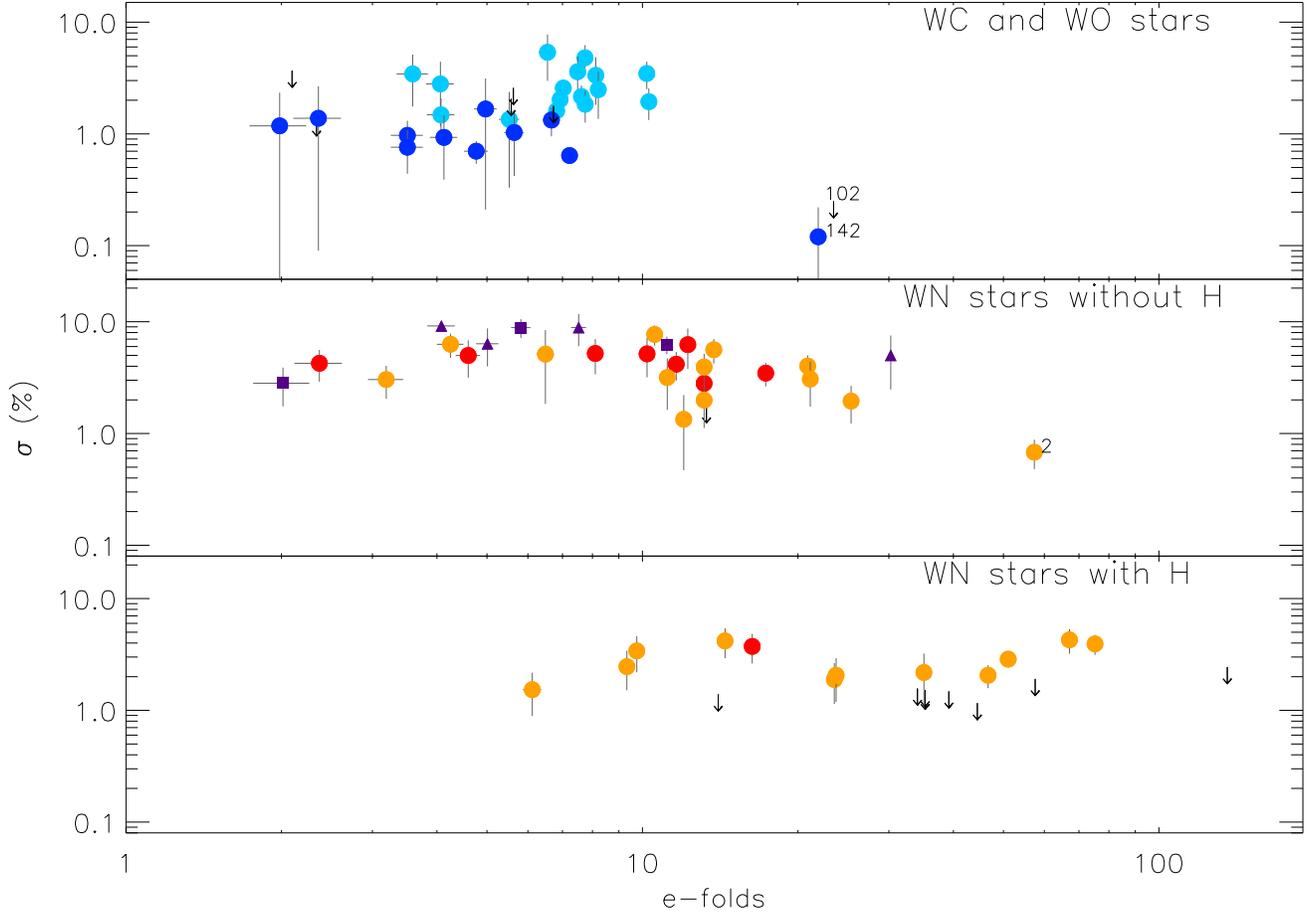}
    \caption{$\sigma$ vs e-folds (see equation\,\ref{efolds}). The spectral subtypes are divided in the same way as in Fig.\,\ref{SigvsT}, and the symbols are the same. However, unlike that figure, all the stars are plotted individually. For clarity, only the stars WR\,2, 102 and 142 are labelled.}
    \label{Sigma_vs_efold}
\end{figure}{}
 
%  x_value_imp[i]=Mdot[pos]*vinf[pos]/(Lum[pos]/3.d8)
%  x_value_imp[i]=70*x_value_imp[i]^(-9/8.)*sqrt(Mass[pos]/40)*sqrt(temp[pos]/40)*(Lum[pos]/(5e5*Lsol))^(-1/4.)
 
\begin{equation}\label{efolds}
    e-folds=70 \sqrt{\frac{M}{40M_\odot}} \sqrt{\frac{T}{40kK}} \left(\frac{L}{5\times10^5L_\odot}\right)^{-1/4} P^{-9/8}
\end{equation}{} 
where P is the wind momentum flow divided by L/c:
\begin{equation}
P=\frac{\dot{M} v_\infty c}{L}
\end{equation}{} 

From that figure, there certainly doesn't appear to be any obvious dependence of sigma on $e-fold$. There are outliers in $e-fold$ which tend to be outliers in $\sigma$ as well, but in the wrong sense to the one required to be explained by LDI. We therefore consider the reduction of LDI with temperature as less likely.

In some sense, there has always been a problem explaining clumping with the LDI alone. The LDI and the $e-fold$ calculation only apply over length scales smaller than the Sobolev length, which is about $R_*(v_D / v_\infty)$ where $v_D$ is the Doppler width of the driving lines, and this is perhaps 1\% of the stellar radius. It is therefore a double-edged sword-- the number of $e-folds$ is high, since the length scale is small, but that produces a lot of tiny features, and lots of them. Increasing the value of P doesn't change the length scale, but does reduce the $e-folds$, which should reduce the value $\sigma$. However, it would probably not be a slow trend and there would be a critical value of P where the instability would to shut off. There is no evidence of any critical P in our data.

\subsection{The Partial Ionization Zone}

If the LDI does not by itself succeed in explaining this data-set (and clumping in general), as proposed at the beginning of this paper, a better accounting for the trends in the data could be given by interpreting the $\sigma$ value as being increased by a SSC zone, which becomes deeper at high $T_*$, as expected. The stars with the weakest-$\sigma$, WR\,2, 102, and 142, could still be consistent with the expectations of a raw LDI without surface stirring by a SSC, 
%[<-- TM: .... except that one does not see growth with time of the lpv, as predicted by LDI!] 
but they can also be regarded as the low-$\sigma$ tail of dying SSC effects. There is also good evidence that the strongest-$\sigma$ stars require some additional process, beyond a SSC: when variations are proven to be periodic, they are related to stellar spots and/or pulsation and the way rotation can organize large-scale surface features into globally clumped features like CIRs.

In that interpretation, convection is necessary to stir the surface, and the non-linearities of line driving yield a visible clumping effect. Convection by itself is not very efficient on wind speed scales, but line driving is notoriously good at producing denser structures out of small inhomogeneities. Furthermore, there is a distinction to make between the LDI, which is a local instability within the wind and only operates on small scales (smaller than the Sobolev length), and the basic line-driven non-linearities, which appear even in the Sobolev approximation (and therefore operate on scales much larger than the Sobolev length). From a basic line-driven non-linearity, any enhancement of the wind density and/or velocity by some 10\% somewhere near the critical point, where the steady mass-loss rate is set would lead to a significantly larger than 10\% effect on the wind density and/or velocity later on farther out in the wind. In other words, surface variations get leveraged by their effect on the local mass flux, and time dependence at the surface breaks the steady-state condition and can produce much larger time dependence in the wind. Therefore, if a SSC is causing clumping, it's not doing so by itself, but that can still be distinguished from the LDI (which might operate on a scale too small to be detectable in optical spectra).

\subsection{Terminal velocity}
Before concluding this section, we wish to point out that our interpretations assume that the $\sigma$ value varies with the surface temperature. \new{But a comparison of the temperatures, $T_\ast$, with the terminal velocities, $v_\infty$, determined by \citet{Ha06} and \citet{Sa12}, shows that they are correlated}. Unfortunately, $v_\infty$ is not determined as accurately as $T_\ast$, yet for higher $T_\ast$, winds are faster. Depending on how the turbulence is generated at the base of the wind and carried away in the wind, one can expect many plausible different density profiles. Because the models currently available fail to describe how parameters \new{commonly used to determine clumping, such as the filling factor, $f$, in atmospheric models, or the vorocity, introduced by \citet{Ow08} and which refers to the porosity in the wind induced by spatial variations in velocity,} depend on $v_\infty$ or $T_\ast$, there is for the moment no clear path for exploring how our observations could be used to test the theory. All we can say for the moment is if the relation of $\sigma$ is with $v_\infty$, it would be in contradiction with any predictions made based on LDI only, as hotter, faster would should in that context show a higher level of clumps-related variability.

\section{Conclusions}

We determined the amplitude of the lpvs for the stars WR\,2, 102 and 142 within 0.1\%. We also revisited the $\sigma$ values (with a $\sim$1\% accuracy) and their relation to the surface temperature, T$_\ast$ (or $v_\infty$) presented in \citet{Mi14}. In addition to confirming higher $\sigma$ values for cooler WR stars (T$\ast \le 50$\,kK), we show that hot WR stars (T$\ast \ge 100$\,kK) show $\sigma$ values much lower than the average. 

Fig.\,\ref{SigvsT} shows an apparent linear relation between $\log(\sigma)$ and T$\ast$. The quality of the data does not allow to determine if that association represents a strict exponential relationship, which could help explore further the nature of the mechanism at the origin of the clumps, neither can we claim the presence of any significant knee in the relationship.

Our preferred theory for the origin of clumps is the presence of a PIZ causing a SSC, which in turn creates perturbations at the base of the wind that are  amplified later by LDI. For higher surface temperature, the PIZ gets closer to the surface. The closer to the surface the SSC is, the lower the energy it can inject at the base of the wind. One could imagine that above a certain temperature, the PIZ is not present and there are no SSC perturbations at the base of the wind. Yet, there is nothing in this study supporting the possibility of this extreme case.

Of course, we recognize that more theoretical work can be required to settle this question. In the meantime, we plan to improve the determination of the dependence of clumping on the surface temperature, and dedicate observing campaigns on monitoring the clumps in WR\,37, 7, 18, 144 and 143. These stars are the very few in the intermediate temperature range between 100 and 120 kK, where the expected $\sigma$ should be around or slightly lower than 1\% of the line amplitude.

\startlongtable
\begin{deluxetable}{r|c|c|c|c|c|c|c|c}
\tablecaption{WC and WO stars \label{tab:WC}}
\tablehead{
\colhead{WR} & \colhead{group} & \colhead{spectral type} & \colhead{T$_{eff}$\tablenotemark{i}} & \colhead{$\sigma$} & \colhead{sp. line} & \colhead{line int.} & \colhead{line FWHM} & \colhead{note}\\
\colhead{} & \colhead{} & \colhead{} & \colhead{(kK)} & \colhead{(\%)} & \colhead{C{\sc iii}/{\sc iv}$\lambda$} & \colhead{(F$_{cont}$)} & \colhead{(km s$^{-1}$)} & \colhead{}
}
\startdata
103 &  (a)  &  WC9d & 45 & 3.34 $\pm$  1.52 & 5696 & 10.6 & 1018 &  \\ 
106 &  (a)  &  WC9d & 45 & 2.49 $\pm$  1.13 & 5696 & 12.4 &  966 &  \\ 
 '' &       &       &    & 3.76 $\pm$  1.58 & 5808 &  3.4 & 1463 &  \\ 
119 &  (a)  &  WC9d & 45 & 3.17 $\pm$  1.27 & 4659 &  7.6 &  932 &  \\ 
 '' &       &       &    & 2.80 $\pm$  1.63 & 5696 & 11.8 &  936 &  \\ 
 '' &       &       &    & 3.26 $\pm$  1.72 & 5808 &  3.1 & 1405 &  \\ 
121 &  (a)  &  WC9d & 45 & 4.97 $\pm$  3.27 & 4659 &  7.5 &  968 &  \\ 
 '' &       &       &    & 4.80 $\pm$  1.42 & 5696 & 11.2 &  936 &  \\ 
 '' &       &       &    & 5.47 $\pm$  1.98 & 5808 &  2.9 & 1434 &  \\ 
81  &  (a)  &  WC9  & 45 & 3.44 $\pm$  1.68 & 5696 & 11.7 & 1099 &  \\ 
 '' &       &       &    & 5.57 $\pm$  1.84 & 5808 &  3.4 & 1396 &  \\ 
92  &  (a)  &  WC9  & 45 & 5.37 $\pm$  2.39 & 5696 &  9.3 &  977 &  \\ 
 '' &       &       &    & 5.91 $\pm$  1.55 & 5808 &  2.6 & 1436 &  \\ 
 '' &       &       &    & 5.32 $\pm$  3.33 & 5880 &  2.9 & 3392 &  \\ 
53  &       &  WC8d & 50 & 3.60 $\pm$  1.31 & 5696 & 11.5 & 1752 &  \\ 
 '' &       &       &    &       $<$   2.92 & 5808 &  5.5 & 1358 &  \\ 
 '' &       &       &    &       $<$   2.92 & 5880 &  2.8 & 2092 &  \\ 
57  &  (b)  &  WC8  & 63 & 3.47 $\pm$  0.96 & 5696 &  5.9 & 2363 &  \\ 
 '' &       &       &    &        $<$  2.49 & 5808 &  6.7 & 1637 &  \\ 
77  &  (b)  &  WC8  & 63 & 2.16 $\pm$  0.55 & 5696 &  4.9 & 2117 &  ii\\ 
 '' &       &       &    & 2.79 $\pm$  0.58 & 5808 &  4.3 & 1875 &  \\ 
135 &  (b)  &  WC8  & 63 & 1.94 $\pm$  0.61 & 5696 &  6.5 & 1988 &  \\ 
 '' &       &       &    & 0.96 $\pm$  0.22 & 5808 &  8.0 & 1434 &  \\ 
14  &  (c)  &  WC7  & 71 & 2.57 $\pm$  0.47 & 5696 &  2.9 & 2852 &  \\ 
 '' &       &       &    & 1.28 $\pm$  0.29 & 5808 &  9.3 & 1917 &  \\ 
50  &  (c)  &  WC7  & 71 & 2.03 $\pm$  0.43 & 5696 &  2.5 & 3745 &  \\ 
 '' &       &       &    & 1.14 $\pm$  0.34 & 5808 &  5.9 & 2394 &  \\ 
90  &  (c)  &  WC7  & 71 & 1.61 $\pm$  0.25 & 5696 &  3.1 & 3094 &  ii\\ 
 '' &       &       &    & 0.68 $\pm$  0.10 & 5808 &  8.1 & 2194 &  \\ 
132 &  (c)  &  WC6  & 71 & 1.48 $\pm$  0.60 & 5696 &  3.4 & 3214 &  \\ 
5   &  (d)  &  WC6  & 79 & 1.33 $\pm$  1.38 & 4659 & 24.3 & 2459 &  iii \\ 
13  &  (d)  &  WC6  & 79 & 1.84 $\pm$  0.58 & 5696 &  2.2 & 3047 &  ii\\ 
27  &  (d)  &  WC6  & 79 & 1.35 $\pm$  1.02 & 5696 &  2.9 & 3048 &  \\ 
 '' &       &       &    & 1.68 $\pm$  0.80 & 5808 & 17.6 & 2086 &  \\ 
45  &  (d)  &  WC6  & 79 & 1.03 $\pm$  0.61 & 4659 & 20.1 & 2432 &  \\ 
154 &  (d)  &  WC6  & 79 & 0.64 $\pm$  1.33 & 4659 & 19.9 & 3019 &  iii \\ 
4   &  (d)  &  WC5  & 79 & 1.67 $\pm$  1.46 & 4659 & 23.5 & 3179 &  iii \\ 
32  &  (d)  &  WC5  & 79 & 0.97 $\pm$  0.34 & 4659 & 20.7 & 2966 &  \\ 
41  &  (d)  &  WC5  & 79 & 0.76 $\pm$  0.32 & 4659 & 17.7 & 4279 &  \\ 
 '' &       &       &    & 1.18 $\pm$  1.09 & 5696 &  2.4 & 4657 &  \\ 
 '' &       &       &    & 0.88 $\pm$  0.32 & 5808 & 21.1 & 3090 &  \\ 
15  &  (e)  &  WC6  & 79 &        $<$  2.58 & 5696 &  2.3 & 4722 &  \\ 
23  &  (e)  &  WC6  & 79 &        $<$  2.06 & 5696 &  2.4 & 3627 &  ii \\ 
17  &  (e)  &  WC5  & 79 &        $<$  1.78 & 5696 &  1.3 & 2689 &  \\ 
33  &  (e)  &  WC5  & 79 &        $<$  1.35 & 5696 & 11.4 & 3750 &  iv \\ 
111 &  (f)  &  WC5  & 89 & 0.70 $\pm$  0.16 & 5808 & 20.2 & 2324 &  \\ 
150 &  (f)  &  WC5  & 89 & 0.93 $\pm$  0.54 & 5808 & 21.4 & 3307 &  \\ 
52  &       &  WC4  & 112&        $<$  3.68 & 5808 & 16.3 & 2436 &  iv \\ 
144 &       &  WC4  & 112& 1.18 $\pm$  1.26 & 6750 &  3.5 & 4294 &  iii \\ 
143 &       &  WC4  & 120& 1.38 $\pm$  1.49 & 4659 &  7.9 & 4602 &  iii \\ 
102 &       &  WO2  & 200&        $<$  0.25 & 5808 &  2.4  &  3539  &  \\ 
142 &       &  WO2  & 200& 0.12 $\pm$  0.10 & 5808 &  1.9  &  3295  &  \\ 
\enddata
\tablenotetext{i}{   From \citet{Sa12}}
\tablenotetext{ii}{   Affected by variable continuum.}
\tablenotetext{iii}{   Affected by continuum normalisation issue.}
\tablenotetext{iv}{   Variations close to the detection threshold.}
\end{deluxetable}

\begin{deluxetable}{r|c|c|c|c|c|c|c|c}
\tablecaption{WN stars without H \label{tab:WN}}
\tablehead{
\colhead{WR} & \colhead{group} & \colhead{spectral type} & \colhead{T$_{eff}$\tablenotemark{i}} & \colhead{$\sigma$} & \colhead{sp. line} & \colhead{line int.} & \colhead{line FWHM} & \colhead{note}\\
\colhead{} & \colhead{} & \colhead{} & \colhead{(kK)} & \colhead{(\%)} & \colhead{He{\sc ii}$\lambda$} & \colhead{(F$_{cont}$)} & \colhead{(km s$^{-1}$)} & \colhead{}
}
\startdata
123 &       &  WN8(WNE-w)  & 44.7  & 5.13 $\pm$ 3.29  & 4686 & 3.6 &  725 & \\
 '' &       &              &       & 5.58 $\pm$ 1.73  & 4860 & 1.8 &  844 & \\
84  &  (a)  &  WN7(WNE-w)  & 50.1  & 6.24 $\pm$ 2.45  & 5411 & 2.4 &  815 & \\
120 &  (a)  &  WN7(WNE-w)  & 50.1  & 6.29 $\pm$ 1.53  & 4686 & 5.1 &  920 & \\
 '' &       &              &       & 6.27 $\pm$ 1.15  & 4860 & 2.0 &  776 & \\
115 &  (a)  &  WN6-w       & 50.1  & 7.70 $\pm$ 1.58  & 4686 & 5.1 & 1149 & \\
 '' &       &              &       & 8.31 $\pm$ 1.89  & 5411 & 1.9 &  967 & \\
55  &  (b)  &  WN7(WNE-w)  & 56.2  & 5.16 $\pm$ 1.98  & 5411 & 2.3 &  858 & \\
67  &  (b)  &  WN6-w       & 56.2  & 4.99 $\pm$ 1.83  & 5411 & 2.3 & 1028 & ii \\
71  &  (b)  &  WN6-w       & 56.2  & 4.17 $\pm$ 1.19  & 5411 & 2.3 & 1114 & \\
75  &  (c)  &  WN6-s       & 63.1  & 4.25 $\pm$ 1.35  & 5411 & 2.2 & 2785 & \\
134 &       &  WN6-s       & 63.1  & 8.86 $\pm$ 1.68  & 4686 &10.6 & 2639 & iii \\
83  &  (c)  &  WN5o        & 63.1  & 2.81 $\pm$ 0.79  & 5411 & 2.1 & 1071 & \\
42d &  (c)  &  WN5b        & 63.1  & 3.94 $\pm$ 1.22  & 4686 & 3.3 & 3390 & \\
20  &  (c)  &  WN5-w       & 63.1  & 4.02 $\pm$ 0.99  & 4686 & 7.4 & 1575 & \\
 '' &       &              &       & 4.52 $\pm$ 1.18  & 5411 & 2.5 & 1233 & \\
34  &  (c)  &  WN5-w       & 63.1  & 5.62 $\pm$ 1.39  & 4686 & 9.0 & 1529 & \\
 '' &       &              &       & 4.99 $\pm$ 2.23  & 5411 & 2.6 & 1324 & \\
54  &  (c)  &  WN5-w       & 63.1  & 3.47 $\pm$ 0.83  & 5411 & 2.2 & 1200 & \\
61  &       &  WN5-w       & 63.1  & 8.89 $\pm$ 2.85  & 5411 & 2.3 & 1243 & iv \\
149 &  (c)  &  WN5-s       & 63.1  & 5.20 $\pm$ 1.81  & 5411 & 2.6 & 1181 & \\
129 &  (c)  &  WN4-w       & 63.1  & 3.07 $\pm$ 1.33  & 4686 & 6.3 & 1639 & \\  
 '' &       &              &       & 3.93 $\pm$ 1.57  & 5411 & 2.4 & 1191 & \\
91  &       &  WN7(WNE-s)  & 70.8  & 9.17 $\pm$ 0.92  & 4686 & 8.2 & 2003 & \\
 '' &       &              &       &10.96 $\pm$ 1.92  & 5411 & 2.4 & 1684 & iv \\
62  &  (d)  &  WN6-s       & 70.8  & 3.04 $\pm$ 0.99  & 4686 & 9.0 & 2353 & \\
 '' &       &              &       & 2.76 $\pm$ 1.11  & 5411 & 2.6 & 2139 & \\
110 &       &  WN5-s       & 70.8  & 2.82 $\pm$ 1.07  & 4686 & 7.4 & 3658 & iii \\
51  &  (d)  &  WN4-w       & 70.8  & 1.95 $\pm$ 0.72  & 4686 & 6.1 & 1648 & \\  
 '' &       &              &       & 2.68 $\pm$ 1.23  & 5411 & 2.1 & 1070 & \\
100 &       &  WN7(WNE-s)  & 79.4  & 6.36 $\pm$ 2.37  & 5411 & 2.2 & 1972 & iv \\
 '' &       &              &       & 6.43 $\pm$ 2.31  & 5876 & 2.1 & 2566 & \\
44  &       &  WN4-w       & 79.4  & 5.01 $\pm$ 2.53  & 5411 & 2.2 & 1286 & iv \\
37  &       &  WN4-s       & 100   & 3.17 $\pm$ 1.54  & 4686 & 9.5 & 3000 & \\
 '' &       &              &       & 3.69 $\pm$ 1.40  & 5411 & 2.5 & 2700 & \\
1   &       &  WN4-s       & 112.2 & 6.24 $\pm$ 1.11  & 4686 & 8.9 & 2615 & \\
 '' &       &              &       & 9.19 $\pm$ 3.52  & 5411 & 2.5 & 2364 & iii \\
7   &       &  WN4-s       & 112.2 & 1.34 $\pm$ 0.87  & 4686 &11.9 & 2059 & \\  
 '' &       &              &       & 2.72 $\pm$ 0.96  & 5411 & 3.1 & 1681 & \\
18  &       &  WN4-s       & 112.2 &  $<$ 1.76        & 5411 & 2.4 & 2916 & ii \\
2   &       &  WN2-w       & 141.3 & 0.60 $\pm$ 0.20  & 4686 & 3.3 & 4294 & \\
\enddata
\tablenotetext{i}{  From \citet{Ha06}}
\tablenotetext{ii}{  Variations close to the detection threshold.}
\tablenotetext{iii}{  Shows CIR-like variations with published periods}
\tablenotetext{iv}{  Shows CIR-like variations without published periods}
\end{deluxetable}

\begin{deluxetable}{r|c|c|c|c|c|c|c|c}
\tablecaption{WN stars with H \label{tab:WNH}}
\tablehead{
\colhead{WR} & \colhead{group} & \colhead{spectral type} & \colhead{T$_{eff}$\tablenotemark{i}} & \colhead{$\sigma$} & \colhead{sp. line} & \colhead{line int.} & \colhead{line FWHM} & \colhead{note}\\
\colhead{} & \colhead{} & \colhead{} & \colhead{(kK)} & \colhead{(\%)} & \colhead{He{\sc ii}$\lambda$} & \colhead{(F$_{cont}$)} & \colhead{(km s$^{-1}$)} & \colhead{}
}
\startdata
79a & (a) & WN9ha        & 35.5 & $<$ 1.52         & 5411 & 1.0 & 140 & \\
79b & (a) & WN9ha        & 35.5 & $<$ 1.44         & 5411 & 1.0 & 101 & \\
108 &     & WN9h         & 39.8 & $<$ 1.48         & 4686 & 2.1 & 381 & ii \\
156 &     & WN8h         & 39.8 &  4.27 $\pm$ 1.06 & 4686 & 3.3 & 419 & \\
124 & (b) & WN8h         & 44.7 &  4.18 $\pm$ 1.25 & 4686 & 3.7 & 689 & \\
 '' &     &              &      &  4.53 $\pm$ 1.84 & 4860 & 2.9 & 849 & \\
158 & (b) & WN7h+Be?     & 44.7 &  3.93 $\pm$ 0.80 & 4686 & 3.2 & 657 & \\
87  &     & WN7h         & 44.7 & $<$ 1.16         & 5411 & 1.2 & 343 & \\
131 &     & WN7h         & 44.7 &  2.18 $\pm$ 1.04 & 4686 & 3.2 & 803 & \\
85  & (c) & WN6h-w(WNL)  & 50.1 &  3.73 $\pm$ 1.10 & 5411 & 2.0 & 728 & ii \\
78  & (c) & WN7h         & 50.1 & $<$ 1.39         & 5411 & 7.6 & 126 & ii \\
20b & (c) & WN6ha        & 50.1 &  2.46 $\pm$ 0.95 & 4686 & 3.3 & 1412 & \\
24  & (d) & WN6ha-w(WNL) & 50.1 & $<$ 1.57         & 5411 & 1.2 & 601 & \\
28  & (d) & WN6(h)-w     & 50.1 & $<$ 1.91         & 5411 & 1.6 & 600 & \\
82  & (e) & WN7(h)       & 56.2 &  3.40 $\pm$ 1.21 & 4686 & 1.2 & 1026 & \\
 '' &     &              &      &  5.13 $\pm$ 0.84 & 5411 & 2.0 & 686 & \\
35  & (e) & WN6h-w       & 56.2 &  1.89 $\pm$ 0.75 & 4686 & 7.0 & 1117 & \\
 '' &     &              &      &  3.91 $\pm$ 1.54 & 5411 & 2.2 & 917 & \\
45c &     & WN5o         & 63.1 &  1.99 $\pm$ 0.87 & 4686 & 7.5 & 1707 & \\
 '' &     &              &      &  3.19 $\pm$ 0.87 & 5411 & 2.3 & 1223 & \\
10  & (f) & WN5ha-w      & 63.1 & $<$ 2.44         & 5411 & 1.5 & 558 & \\
136 & (g) & WN6(h)-s     & 70.8 &  1.53 $\pm$ 0.64 & 4686 &12.0 & 2027 & \\
 '' &     &              &      &  4.61 $\pm$ 1.02 & 5411 & 2.3 & 1940 & \\
 '' &     &              &      &  6.13 $\pm$ 0.95 & 5876 & 1.6 & 2438 & \\
128 & (g) & WN4(h)-w     & 70.8 &  2.06 $\pm$ 0.87 & 4686 & 4.3 & 1742 & \\
 '' &     &              &      &  2.10 $\pm$ 0.55 & 4860 & 1.7 & 949 & \\
 '' &     &              &      &  3.05 $\pm$ 0.75 & 5411 & 1.7 & 918 & \\
152 &     & WN3(h)-w     & 79.4 &  2.87 $\pm$ 0.46 & 4686 & 4.1 & 1772 & \\
3   &     & WN3h-w       & 89.1 &  2.06 $\pm$ 0.48 & 4686 & 2.4 & 1742 & \\
\enddata
\tablenotetext{i}{   From \citet{Ha06}}
\tablenotetext{ii}{   Variations close to the detection threshold.}
\end{deluxetable}

%\startlongtable
\begin{deluxetable}{r|c|c|c|c|c|c|c|c}
\tablecaption{WN/C stars \label{tab:WNC}}
\tablehead{
\colhead{WR} & \colhead{group} & \colhead{spectral type} & \colhead{T$_{eff}$\tablenotemark{i}} & \colhead{$\sigma$} & \colhead{sp. line} & \colhead{line int.} & \colhead{line FWHM} & \colhead{note}\\
\colhead{} & \colhead{} & \colhead{} & \colhead{(kK)} & \colhead{(\%)} & \colhead{He{\sc ii}$\lambda$} & \colhead{(F$_{cont}$)} & \colhead{(km s$^{-1}$)} & \colhead{}
}
\startdata
88  &  & WC9/WN8 & 40 &  5.81 $\pm$ 1.66 & 5696 & 7.7 & 1018 & \\
 '' &  &         &    &  6.23 $\pm$ 1.56 & 5808 & 2.9 & 1157 & \\
 '' &  &         &    &  4.58 $\pm$ 2.24 & 5880 & 2.9 &  946 & \\
126 &  & WC5/WN  & 63 &  1.72 $\pm$ 0.55 & 4686 & 3.1 & 2734 & \\
 '' &  &         &    &  1.18 $\pm$ 0.34 & 5808 & 6.4 & 2310 & \\
8   &  & WN7o/CE & 79 &  5.68 $\pm$ 1.47 & 5411 & 2.1 & 1115 & \\
 '' &  &         &    &  4.47 $\pm$ 0.96 & 5808 & 4.0 & 1557 & \\
26  &  & WN7/WCE & 79 &  4.98 $\pm$ 0.76 & 5411 & 2.3 & 3005 & \\
58  &  & WN4/WCE & 79 &  4.05 $\pm$ 1.95 & 5411 & 3.0 & 1800 & \\
\enddata
\tablenotetext{i}{From \citet{Ha06} and \citet{Sa12}}
%\tablecomments{Note that {\tt \string \colnumbers} does not work with the  vertical line alignment token. If you want vertical lines in the headers you can not use this command at this time.}
\end{deluxetable}

\newpage

\acknowledgments

ANC gratefully acknowledges support from the Gemini Observatory, which is operated by the Association of Universities for Research in Astronomy (AURA) under a cooperative agreement with the National Science Foundation, on behalf of the Gemini Observatory partnership: the National Science Foundation (United States), National Research Council (Canada), Agencia Nacional de Investigaci\'{o}n y Desarrollo (Chile), Ministerio de Ciencia, Tecnolog\'{i}a e Innovaci\'{o}n (Argentina), Minist\'{e}rio da Ci\^{e}ncia, Tecnologia, Inova\c{c}\~{o}es e Comunica\c{c}\~{o}es (Brazil), and Korea Astronomy and Space Science Institute (Republic of Korea).
AFJM and NSL thank NSERC (Canada) for financial aid.
Based on observations obtained at the Gemini Observatory, processed using the Gemini IRAF package. The program ID are GS-2008B-Q-87, GS-2010B-Q-58, GN-2010B-Q-78, GN-2010B-Q-119, GN-2012B-Q-115, GS-2014A-Q-42, GS-2014A-Q-73, GN-2017B-Q-6, GS-2017B-Q-62 and GN-2018B-Q-404.


\vspace{5mm}
\facilities{Gemini(GMOS), OMM(Spectrograph), CTIO:1.5m}


\software{IDL Astronomy Library \citep{La93}
%          astropy \citep{2013A&A...558A..33A}%,  
%          Cloudy \citep{2013RMxAA..49..137F}, 
%          SExtractor \citep{1996A&AS..117..393B}
          }

\end{document}